\documentclass[12pt]{iopart}
\usepackage{graphicx}
\usepackage[utf8]{inputenc}
\usepackage{bm}
\usepackage{color}
\usepackage{subfig}
\usepackage{amssymb}
\begin{document}

\title[Local disalignment can promote coherent collective
motion]{Local disalignment can promote coherent collective motion
  through rapid information transfer}

\author{M. Meschede$^1$ and O. Hallatschek$^2$}
\address{Biophysics and Evolutionary Dynamics Group, Max Planck Institute for Dynamics and Self-organization, 37073 G\"ottingen, Germany}
\ead{\mailto{matthias.meschede@googlemail.com$^1$}, \mailto{oskar.hallatschek@ds.mpg.de$^2$}}

\begin{abstract}
  When particles move at a constant speed and have the tendency to
  align their directions of motion, ordered large scale movement can
  emerge despite significant levels of noise. Many variants of this
  model of self-propelled particles have been studied to explain the
  coherent motion of groups of birds, fish or microbes. Here, we
  generalize the exactly aligning collision rule of the classical
  model of self-propelled particles to the case where particles after
  the collision tend to move in slightly different directions away
  from each other, as characterized by a collision angle $\alpha$. We
  map out the resulting phase diagram and find that, in sufficiently
  dense systems, small disalignment can lead to higher global
  alignment of particle movement directions. We show that in this
  dense regime, global alignment is accompanied by a grid-like spatial
  structure which allows information to rapidly percolate accross the
  system by a ``domino'' effect. Our results
  elucidate the relevance of disalignment for the emergence of
  collective motion in models with exclusively repulsive interaction
  terms.
\end{abstract}

\maketitle

%=========================================================================================
\section{Introduction}
\begin{figure}[htb!]
\center{  \subfloat[Vicsék]{\label{fig:Vicsek1}\includegraphics[width=0.3\columnwidth]{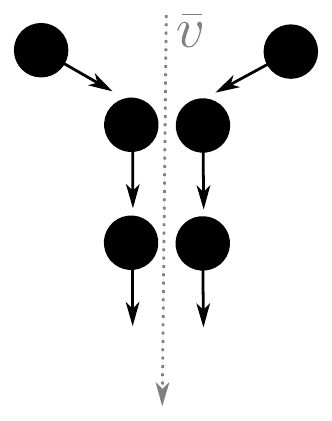}}\hspace{0.09\columnwidth}
  \subfloat[Disalignment]{\label{fig:disalign}\includegraphics[width=0.3\columnwidth]{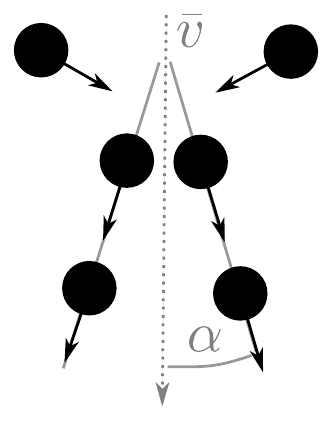}}}
  \caption{Self-propelled particle models describe the coherent motion of
particles that move at a constant velocity and interact when they collide. In
the classical model (a) due to Vicsek et al.~\cite{Vicsek1995}, colliding particles align
their direction of motion. The direction after the collision is obtained from
averaging the velocities prior to the collision. b) Here, we study a more general
model in which the velocity directions after the collision deviate by an angle
$\alpha$ from the averaged direction. The case $\alpha = 0$ leads to the original
Vicsek model.}
  \label{fig:Fig1cartoon}
\end{figure}

\begin{figure}[htb!]
\center{  \subfloat[$\alpha = 0^\circ$]{\label{fig:snapvicsek}\includegraphics[width=0.3\columnwidth]{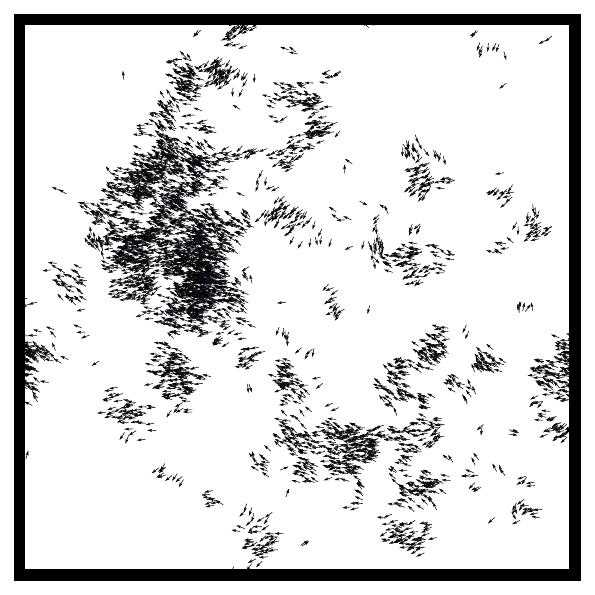}}
  \subfloat[$\alpha=1^\circ$]
{\label{fig:snapdisalign1}\includegraphics[width=0.3\columnwidth]{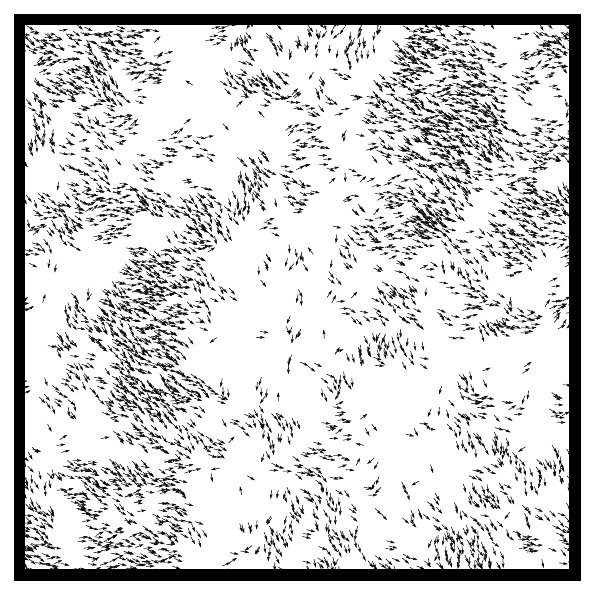}}
  \subfloat[$\alpha=10^\circ$]
{\label{fig:snapdisalign}\includegraphics[width=0.3\columnwidth]{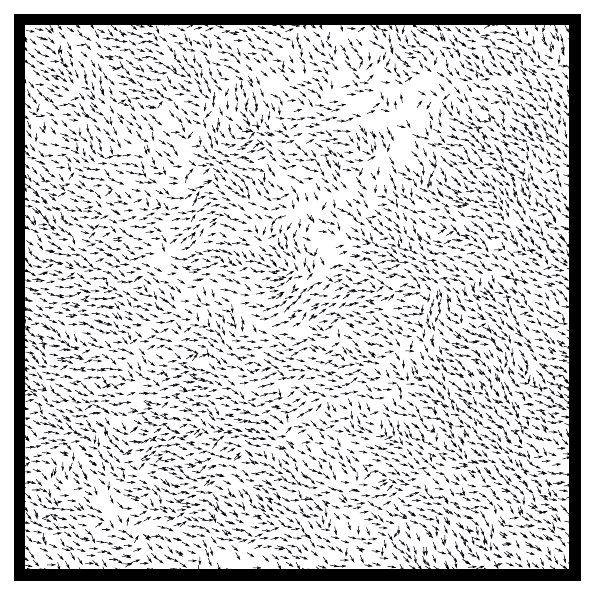}}}
\caption{Characteristic snaphots of the distribution of particle
  positions and movement directions (arrows) in our disalignment
  model. (a) For vanishing disalignment angle $\alpha$, the original
  Vicsék model is obtained with its characteristic flocking
  structure. As the disalignment angle is increased (b, c), the
  distribution of particles changes markedly - density fluctuations
  are suppressed. As we argue in this article, the resulting homogeneous
  distribution of particles leads to a change in the mechanism that
  drives global order. (Other simulation parameters were $N=4096$,
  $\eta=25^\circ$) }
  \label{fig:snapshots}
\end{figure}

The emergence of ordered motion in groups of interacting particles
that move at a constant speed is reminiscent of the collective motion
observed in many animate and inanimate systems~\cite{Vicsek2012}. A
wide variety of different models of such self-propelled particles
(SPPs) have been explored with the goal to quantify conditions for the
global alignment of the movement directions of the individual
particles. These models have in common that they rely on local
interaction rules, as they are thought to apply to many animal
swarms~\cite{Vicsek2012,Gregoire2001,Gregoire2003,Buhl2006,Chate2008a,Couzin2002,Couzin2003},
and that the movement directions of particles are continually
perturbed by random noise. 

The first and most basic of these models of SPPs is due to Vicsék et
al.~\cite{Vicsek1995}, and relies on an explicit alignment interaction
that adjusts each particle's movement direction to the average
direction of its surrounding particles, see
Fig. \ref{fig:Vicsek1}. For sufficiently weak noise levels,
self-organized collective motion results from the local interaction
rule that the movement directions of colliding individuals are
aligned. For strong noise levels, the system will inevitably fail to
order globally. The amount of coherent collective motion can be
measured by an 'order parameter' defined as the magnitude of the
globally averaged velocity vector.  While an analog of an equilibrium phase
transition is obtained in the limit of zero velocities, the
order-to-disorder transition is generally a unique non-equilibrium
phenomenon as it is driven by the perpetual motion of the interacting
entities~\cite{Vicsek1995}. Whether the
transition between the two states is continuous or discontinuous (in the
thermodynamic limit) has been intensely
debated~\cite{Vicsek2012,Chate2008b,Aldana2007,Nagy2007}. Similar phase
transitions are observed in variants of the classical Vicsék model that add
cohesive and a repulsive interaction
terms~\cite{Vicsek2012,Couzin2002,Gregoire2004,Chate2008a}.

Contrary to the Vicsék model and its variants, a second group of SPP models
\cite{Grossman2008} does not introduce an explicit alignment but only
an isotropic repulsive force, repelling nearby particles. Surprisingly
an ordered phase can be observed even then: The perpetual motion of
the SPPs leads to a weak alignment through each collision and, when
enough collisions are accumulated, order emerges given weak enough
noise~\cite{Grossman2008,Vicsek2012}.

The observation of local alignment causing global alignment, replicated many
times, suggests that higher local alignment will always lead to stronger
ordering. Furthermore one might think that the Vicsék model, perfectly aligning
the SPPs locally, exhibits the highest levels of global order among all models
with the same noise strength, particle density and interaction range.

We demonstrate in the following that, contrary to this intuition,
local disalignment can even enhance global order. To show this, we
generalize the Vicsék collision rule such that the velocity vectors
after collision diverge by a small angle $\alpha$. The disalignment
angle $\alpha$ is chosen to point away from the center of mass of the
interacting particles, which results in a repulsive interaction, see
Fig \ref{fig:disalign}. For any finite disalignment angle, our
interaction rule tends to reduce local order compared to the classical
Vicsék model. Yet, our numerically determined phase diagram shows that
global order can be increased for a nonzero disalignment, for certain
densities and noise levels. We argue that this effect is ultimately
the manifestation of reduced density fluctuations in the presenced of
disalignment, which leads to a more efficient information transfer
across the system.

%=========================================================================================

\section{The disalignment model}
Our model is a generalization of the classical Vicsék
model~\cite{Vicsek1995}. In two dimensions, the orientation of each particle can
be characterized by an angle $\Phi$, measured in the
counter-clockwise direction.  In each timestep, the angle $\Phi_i$
corresponding to a focal particle $i$ is updated according to the
rule
\begin{eqnarray} 
\Phi_i(t+\Delta t) = \overline{\Phi}^{(r)}_i(t)~\pm \alpha~+\Delta \Phi \;,
\label{eq:dir_disalignment}
\end{eqnarray}
as illustrated in Fig. \ref{fig:disalign}. Here, the angle
$\overline{\Phi}^{(r)}_i(t)$ characterizes the average orientation of
all particles within a circle of radius~$r$ centered at the focal
particle. The random noise term $\Delta \Phi$ is chosen uniformly from
the interval $[-\eta,\eta]$. Both, the level of noise $\eta$ and the
disalignment angle $0<\alpha<180^\circ$ are measured in arcdegrees in
the following. All angles are measured in the counter-clockwise
direction. Finally, the parameter $\alpha\geq0$ is the disalignment
angle -- the new parameter of our model. The sign in front of $\alpha$
is always chosen such that the particles tends to move away from the
line that projects from the average position along the average
movement direction of all particles inside the interaction
radius~\footnote{For definiteness, we set $\alpha = 0$ when a particle
  is alone within its interaction range.}. As a consequence, the
$\alpha$ term drives particles away from one another, as illustrated
in Fig.~\ref{fig:Vicsek1}. Note that the original Vicsék model is
obtained when the disalignment angle $\alpha$ is set to zero. The
parameter $\alpha$ can also be viewed as a tuning wheel by which the
exact alignment interaction can be
broken systematically.  % (See Appendix~A for details of the computer
% implementation.)

The position $\vec x_i$ of particle $i$ is consequently updated as:
\begin{eqnarray} 
\vec{x}_i(t+\Delta t) = \vec{x}_i(t) + v
\left(
{\cos(\Phi_i(t+\Delta t))\atop
\sin(\Phi_i(t+\Delta t))}
\right)
\Delta t
\label{eq:pos_disalignment}
\end{eqnarray}
Throghout the reported simulation results, we chose the magnitude of the particle velocity constant as~$v=0.1$, the
timestep~$\Delta t = 1$ and the interaction radius~$r=1.0$. The~$N$~particles
move in a square cell with periodic boundaries of length~$L$. The particle
density is given by~$\rho = N/L^2$.

We measure the degree of global alignment by the order parameter~$\varphi$,
\begin{eqnarray}
\varphi = \frac{1}{|v|}|\sum_i^N \vec{v}_i| \;,
\end{eqnarray}
which represents the normalized average particle velocity.
%==================================================================
\begin{figure}[htb!]
 \ \subfloat[]{\label{fig:phase:a}\includegraphics{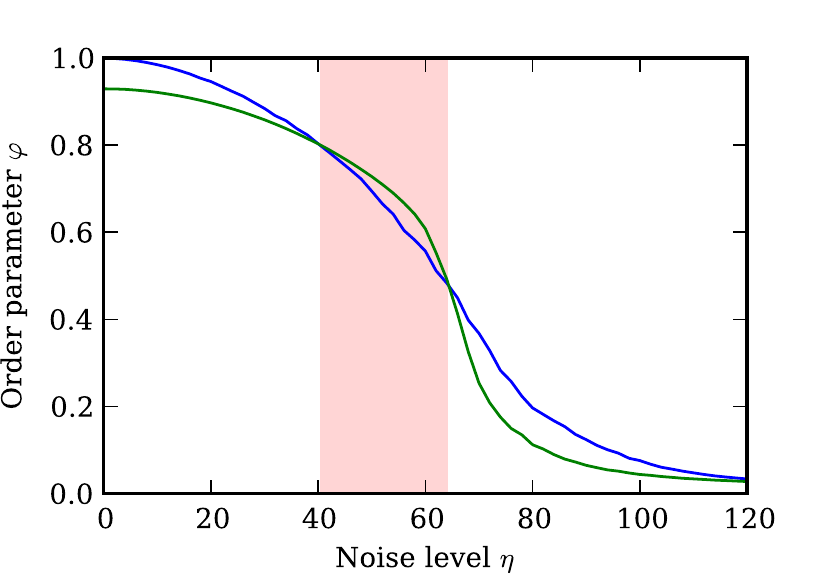}}
  \subfloat[]{\label{fig:phase:b}\includegraphics{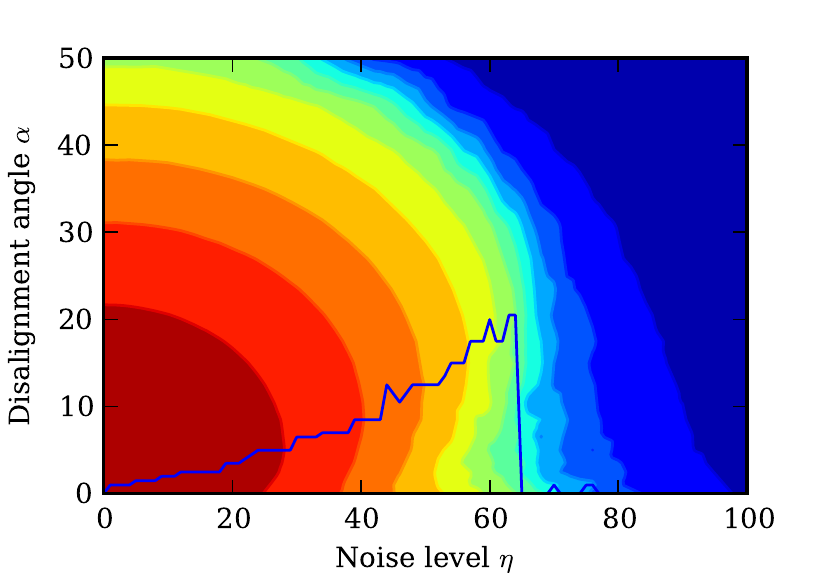}}\\
  \subfloat[]{\label{fig:phase:c}\includegraphics{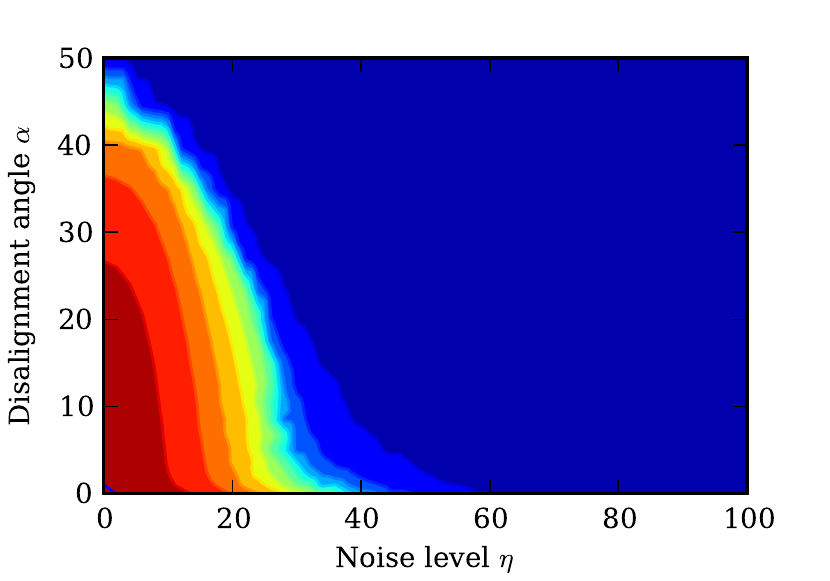}}
  \subfloat[]{\label{fig:phase:d}\includegraphics{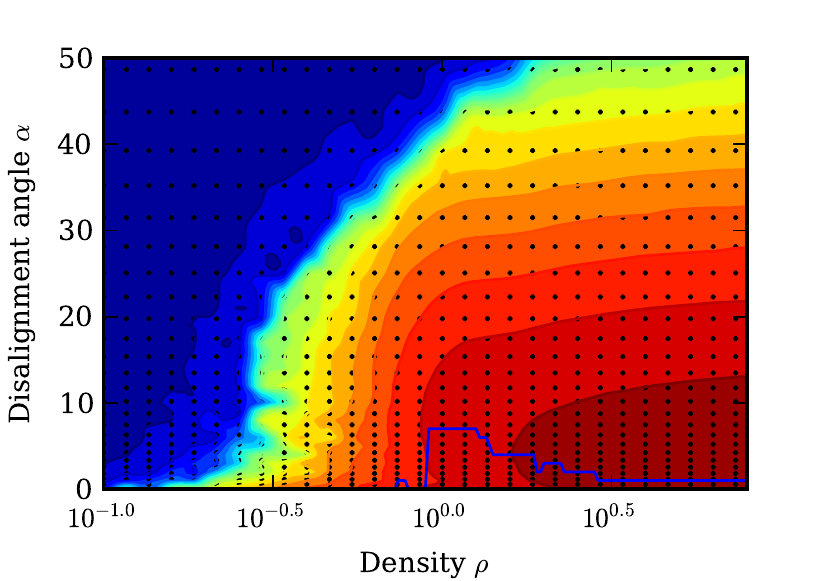}}
  \caption{Phase diagrams summarize the global orientational order in our
    disalignment model, and show that collective motion can be
    promoted by a small degree of disalignment. a) Order parameter
    $\varphi$ as a function of noise $\eta$ in the classical Vicsék
    model ($\alpha = 0^\circ$, blue) and the disalignment model
    ($\alpha = 20^\circ$, green), respectively, at a particle density
    of $\rho=2$. For medium noise levels (same density), the
    disalignment model has a higher order parameter and exhibits a
    sharper transition to the disordered phase (shaded region). In the
    heat plots b), c) and d), the order parameter $\varphi$ is
    indicated by color (red=1,blue=0) as a function of disalignment
    angle~$\alpha$ and noise~$\eta$ (b,c) and as a function of of
    $\alpha$ and $\rho$ (d), respectively. Figs. b,c) differ in their
    densities (b:$\rho=2$, c: $\rho=0.5$). In Fig. d), the noise level
    was fixed at $\eta=20^\circ$. The optimal disalignment angle
    (maximum in vertical direction) is indicated by the blue
    line. Measurement points are shown in d) as black dots. Other
    parameters were $N=2048$ and $v=0.1$.}
  \label{fig:phase2d}
\end{figure}

\section{Snapshots and phase diagrams}
After randomizing the inital particle positions and orientations, the
particles start to move and interact through collisions. For small disalignment
$\alpha$, each collision tends to align the colliding particles and, as in the
original Vicsék model, dense groups of aligned particles form moving jointly through the system~\cite{Vicsek1995}, see Fig.~\ref{fig:snapvicsek}.
The abundance of particles within an interaction range inside a dense group
reduces the effect of noise by averaging the movement directions of many
particles. Depending on the noise strength and density, these groups can further
align among each other, thus leading to a nonzero order parameter $\varphi$. The
smaller the number and size of the groups, the less frequent are the
interactions between them.

As one increases $\alpha$ to $1^\circ$, the formation of dense groups is suppressed
due to the repulsive effect of the disalignment interaction.  More
loosly connected groups form instead, which occupy a much larger area
of our simulation box. On the one hand, this leads to more frequent collisions between
clusters, as illustrated by the snapshot in
Fig.~\ref{fig:snapdisalign1}. On the other hand, the number of particles within an
interaction range decreases inside clusters, thus amplifying the effect of
noise. 

If we further increase $\alpha$ to $10^\circ$, dense groups become
very rare and a grid-like structure forms that spans most of the
simulation area (Fig.~\ref{fig:snapdisalign}). Such a system spanning
grid can, however, only form for large enough mean densities, $\rho
\gtrsim 1$. For very low densities, the repulsive interaction
disintegrates any cluster of particles such that solitary particles
move in random directions.

The phase diagrams in Fig.~\ref{fig:phase2d} summarize the behavior of our
model. The first plot (Fig.~\ref{fig:phase:a}) shows the behavior of the order
parameter $\varphi$ as a function of noise level $\eta$ for the original Vicsék
model with $\alpha=0$ and for finite disalignment with angle $\alpha=20^\circ$.
Both systems are highly ordered for small noise and become disordered as noise
levels are increased. At zero noise, the Vicsék model approaches perfect global
order with $\varphi = 1$ while the disalignment model retains less order, as
one might expect. At medium noise levels, however, the disalignment rule for
$\alpha=20^\circ$ leads to higher global order than for $\alpha=0^\circ$. This
counterintuitive behavior at intermediate noise levels is the focus of our
discussion below. Also note that the transition from the ordered to the
disordered phase appears to be sharper for disalignment than for alignment.

The heat plot in Fig.~\ref{fig:phase:b} shows a two-dimensional phase
diagram, in which the order parameter is indicated as a function of
both the noise level and the disalignment angle, for a similar density
as in Fig.~\ref{fig:phase:a}.  Again, the asymptotic behavior follows
intuition: high noise levels and large disalignment angles together
prevent order. Highest order is achieved for zero noise levels and
zero disalignment angles. The intermediate behavior, however, again
shows the surprising phenomenon of an 'optimal' disalignment angle
that leads to the highest order for a given noise level $\eta$. This
angle is indicated as a blue line and increases with increasing noise
levels.

Disalignment can only promote global alignment when densities are of
order $\approx 1$ or larger. For lower densities, the order parameter is
always largest for vanishing disalignment, as can be appreciated from
Fig.~\ref{fig:phase:c} $\rho=0.5$. At these densities, smaller
noise levels suffice to break down order, even more so for finite
disalignment angles. 

The heat plot in Fig.~\ref{fig:phase:d} finally depicts the dependence
of the order on both the disalignment angle and the particle density,
for a fixed level of noise. As one increases the density of the
system, an optimal disalignment angle appears at $\rho\approx 1$, and
decreases as one further increases the density of the system. The
sharp contrast of the optimal angle at $\rho$ greater or less than
$\approx 1$ and the different behavior indicated in the snapshots in
Fig.~~\ref{fig:snapvicsek} suggests that the spatial distribution of
the particles, in dense groups or a grid-like structure, could play a
crucial role for explaining our results.

\section{Information transfer}

\begin{figure}
  \subfloat[Original Vicsék model]{\label{fig:infvicsek}\includegraphics{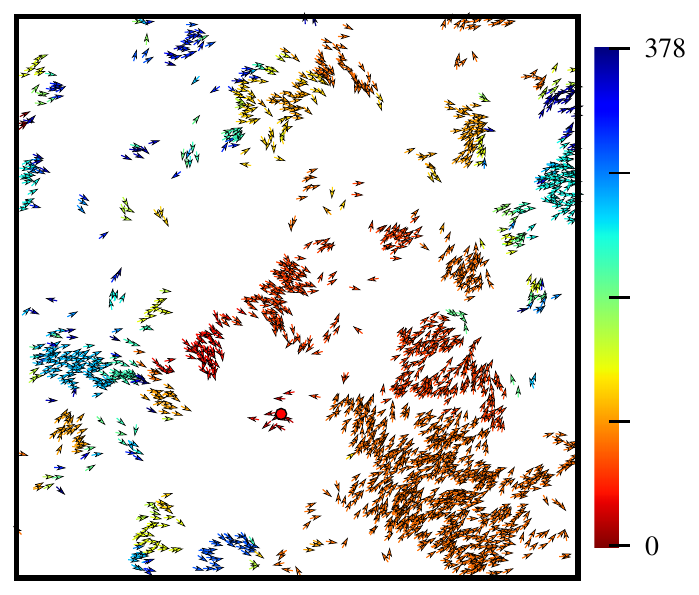}}
  \hspace{0.05\textwidth}
  \subfloat[Disalignment model]
{\label{fig:infdisalign}\includegraphics{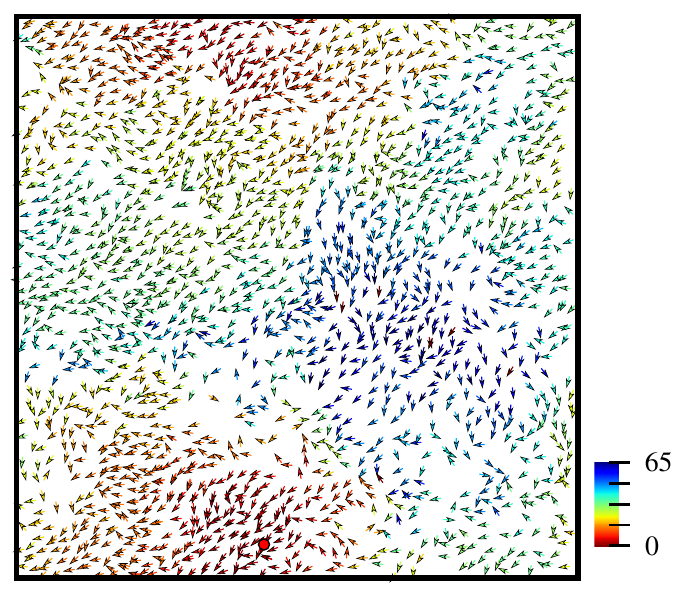}}\\
  \subfloat[Distribution of influence times]
{\label{fig:distribution}\includegraphics{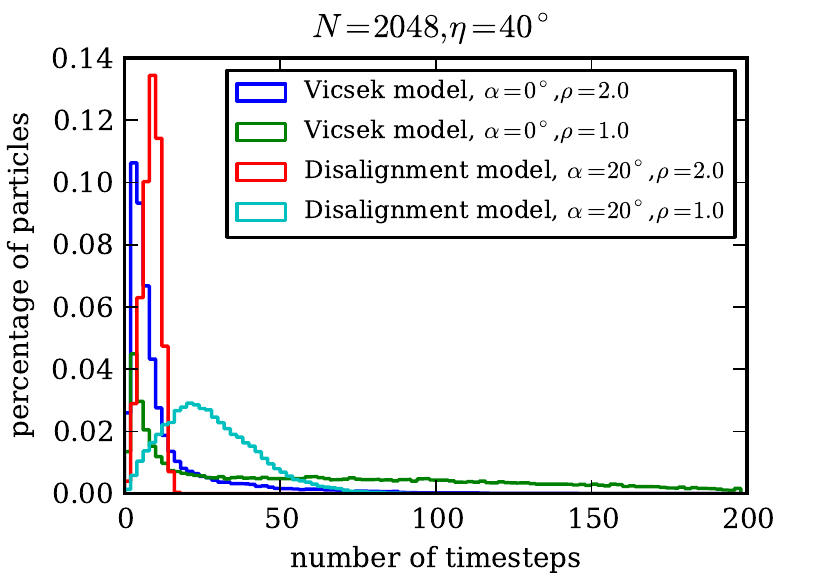}}
  \subfloat[Maximal influence time]
{\label{fig:inftime}\includegraphics{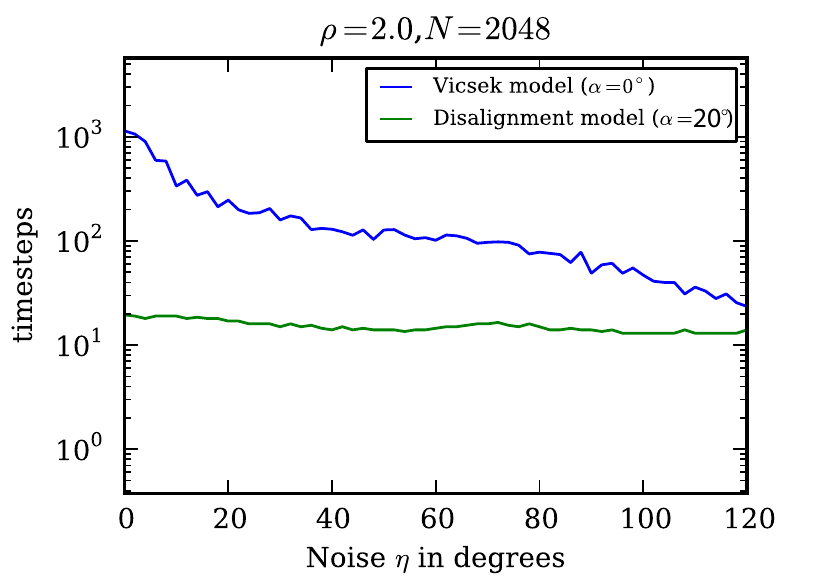}}
\caption{Information spreads differently with and without
  disalignment. Figs. a, b) show snapshots of particle
  positions and movement directions (arrows) at a given point in
  time. The color of a particle codes for the additional delay time
  until it is ``influenced'' (as defined in the main
  text) by the current state of a
  focal particle (red dot). In Fig. a), original Vicsék model
  ($\rho=1.0$, $N=2048$, $\eta=30^\circ$), influence spreads in chunks
  from group to group.  In b), disalignment model ($\rho=1.0$,
  $N=2048$, $\eta=30^\circ$, $\alpha=10^\circ$), influence spreads
  evenly and much faster through the system, which is spatially
  organized in a grid structure. Fig. c) shows distributions of
  influence times for the Vicsék and the disalignment model
  ($\alpha=20^\circ$) at two different densities and equal noise
  ($N=2048$, $\eta=40^\circ$).  For both densities, the distribution
  corresponding to the Vicsék model exhibits a long time tail and a
  peak, which is due to interactions within and between ``flocks'', as
  we argue in the main text. These features are absent in the
  grid-structured regime of the disalignment model, where information
  spreads through a different mechanism. Fig. d) shows the maximal
  influence times for different noise levels and many
  realizations. Notice that disalignment (with $\alpha=20^\circ$) can
  reduce these maximal influence times by orders of magnitudes.}
  \label{fig:information_time}
\end{figure}

\subsection{The role of information transfer}
In models of collective motion, global coherent order emerges from an
interplay between aligning and random disaligning forces. The aligning
forces allow orientational information to be transmitted from one
particle to its close neighbours. The global effect of these driving
forces on order does not only depend upon the degree of alignment in
an interaction. It also depends to a large extend on the ability of
the system to exchange information about movement directions between
all particles: A particle that never interacts with a group of aligned
particles will never align with them, no matter how strong the
aligning force might be. On this view, low density therefore should
generally decrease the order of a system because the number of
interactions between particles is reduced such that information about
the orientation of a particle travels more slowly through the system while
noise deteriorates the information during the transmission. This
behavior has often been observed in previous studies~\cite{Vicsek1995,
  Baglietto2008, Baglietto2009} and also characterizes our model, see
Fig.~\ref{fig:phase:d}.

% CHANGE OR ERASE: In the Vicsék model, noise can even be
% rescaled to a universal noise, that leads to the same critical value
% for all densities according to $\eta^* = \eta \rho^\kappa$ with
% $\kappa =0.5$ \cite{Baglietto2009,Czirok1997}. In our model, such a
% rescaling is not as easily possible because
% it not only stretches the phase transition along the noise axis as can
% be seen by comparing the two curves in Fig.~\ref{fig:phase:a}.  It
% might however be possible to find a two dimensional mapping involving
% disalignment angle as well as noise that projects two dimensional
% phase diagrams of different densities, e.g. phase diagram
% Fig.~\ref{fig:phase:b} and Fig.~\ref{fig:phase:c}, onto each other.
% STOP

% CHANGE OR ERASE: Apart from the strength of alignment in the interaction term, a further useful
% quantity to compare the influence of aligning and disaligning forces in the
% system is the rate of a particle's interactions with other particles with the
% time it takes to randomize its movement direction due to noise.  The
% interaction rate can be defined differently, e.g. as the number of interactions
% with any particle, the number of interactions with the whole system and many
% other methods. Since information about alignment is transmitted via interactions,
% this rate will be connected to global order.
% STOP
 
In models of collection motion, information about movement directions
spreads in two distinct ways:

i) Neighboring particles closer than the interaction distance directly
adjust their movement directions with respect to one
another. The orientation of particles within a connected clusters can be aligned after a few
timesteps by this direct interaction through a 'domino' effect: First
neighbours within an interaction range align and encode the
information about the directions of their neighbors in their own new
direction. In the next step, they propagate this information to their
nearest neighbours such that information spreads until there are no
more new particles within an interaction range of the considered
cluster of particles. While the positions of all particles change
during this process, the information transmission does not rely on
particle movement (in contrast to the second mechanism of information
transmission, below) - it is also present in the limit of zero
velocities. 

% Directions of all particles continuously change due to noise
% and other encounters with other particles such that the encoded
% orientational information randomizes over time and distance. However,
% every particle receives information from many neighbours through many
% different paths stabilizing the noisy transmission.

ii) The second type of information flow in models of collective motion depends on
the movement of particles. This property fundamentally distinguishes
self-propelled particle models from equilibrium models. A particle can move to
a distant location and exchange orientational information with a particle there.
Or in the context of the whole system, a particle can be used as a 'messenger'
that exchanges orientational information between two or more other particles.  In
contrast to the above described direct interaction, this mechanism works
over large distances even without a continuous chain of neighbours that are
separated by at most one interaction distance. However, since information
deteriorates by noise while a particle is on its way, single particles
can transmit information only over short distances. Therefore, dense groups of
interacting particles, or ``flocks'', play an important role for preserving orientational
information against randomization through noise: A group continually
averages the orientations of all of its members and can thus retain its average
velocity over longer times and travel distances. A clustered structure as
in the original Vicsék model is therefore crucial for this type of long
distance information transfer \cite{Chate2008b, Nagy2007, Wu2000,
Gregoire2001}). 

Information transmission through the directed motion of self-propelled
particles is suppressed in the presence of a disalignment rule because
the formation of groups is hampered by the repulsive character of the
interaction.  As a consequence, order breaks down below a certain
density when there are not enough neighbours to ensure a continuous
chain of information transfer. 

For large enough densities , however, a grid-like structure observed at density $\rho>1$ in the
disalignment case allows the whole system to receive orientational
information through this mechanism. Information then percolates
through the system in a similar manner as in the equilibrium ``XY''
spin model. For the XY-model, however, it can be theoretically shown
that full order can not be achieved in 2 dimensions such that an SPP
model certainly requires movement of particles to achieve an ordered
state far from equilibrium~\cite{mermin1966absence,toner2005hydrodynamics}.

For large enough densities ($\rho>1$), however, a grid-like structure forms in the
disalignment case, which allows the whole system to receive orientational
information through the first ``domino'' mechanism. Information then percolates
through the system in a similar manner as in the equilibrium ``XY''
spin model. For the XY-model, however, it can be theoretically shown
that full order can not be achieved in 2 dimensions such that an SPP
model certainly requires movement of particles to achieve an ordered
state far from equilibrium~\cite{mermin1966absence,toner2005hydrodynamics}.

We expect (and test below) that this domino mechanism of information
transfer is typically much faster than information transfer by moving
flocks, simply because the signal propagation is independent of the
motion of the signal carriers.

\subsection{The maximum speed of information transfer}
We now demonstrate that information can indeed be transmitted by two
different mechanisms at very different speeds depending on the chosen
model parameters. To this end, we measured how quickly a particle
interacts with all other particles through a chain of succesive
interactions. This can be done, by choosing a focal particle and
following through time how other particles became influenced by this
particle.  Operationally, we define the set of 'influenced' particles
as follows: i) At time $t_0$ no particles is influenced by the focal
particle, yet. ii) At $t>t_0$, particles become influenced by the
focal particle if they either interact directly with the focal
particle, or indirectly via an already influenced particle. The number
of influenced particles increases over time and measures an upper
limit for the speed at which the focal particle is able to transmit
any information to other particles. In Fig.~\ref{fig:information_time}
we display the time until particles are influenced by a randomly
chosen focal particle, marked by a red dot, for a) the Vicsék model
and b) the Disalignment model with $\alpha = 10^\circ$ in a single
realization. Particle positions are shown at time $t_0$. Notice that
the influence of the focal particle spreads much more quickly in the
presence of disalignment. Furthermore, information spreads in chunks
from group to group in the Vicsék model. The disalignment model, on
the other hand, circulates orientational information of the focal
particle evenly throughout the system and reaches the last particles
much faster. Note our measure of the speed of information transfer
does not entail any statement about the quality of the transmitted
information, which deteriorates through the continual perturbation by
noise. Nevertheless, it demonstrates the qualitative difference
between the two types of information transfer that we described above.

Fig.~\ref{fig:distribution}) shows the measured distributions of
transmission times obtained from many runs by averaging over the
choice of the focal and influenced particle. Note that the difference
between the case with disalignment and with strict alignment are
particularly prominent at $\rho\approx 1$, where the Vicsék model
exhibits a broad class of particles that need many time steps to
receive any information from a focal particle. The distribution is
peaked at the characteristic size of particle groups, which exchange
information on a fast time scale. In Fig.~\ref{fig:inftime}, we report
the time until the last particle of the system is influenced by a
focal particle for many realizations. This maximum transmission time
is, for low noise, orders of magnitude larger than in the disalignment
model. Interestingly, neither of the models displays a sudden change
of the maximum influence time at the critical noise level where global
order breaks down.

\section{Conclusions}

% Here? Considering snapshots and phase diagrams of the previous section, we
% expect the domino mechanism predominantly responsible for alignment
% for finite disalignment angle at sufficiently high density, and the
% second mechanism for vanishing disalignment angle, which corresponds
% to the original Vicsék model.

We have demonstrated that changing the strictly aligning interaction
of the classical Vicsék model to a slightly disaligning interaction
can lead to an increase in global order.  The beneficial effect of the
disalignment term on global order is most prominent for densities
close to $\sim 1$ when the spatial structure of the population was
found to be fundamentally changed by the repulsive disalignment
interaction. Isolated groups or flocks, as found in the classical
Vicsék model, were instabil and disintegrated in the presence of the
disaligning interaction. As a consequence, a homgeneous grid-like
spatial structure formed that spaned the whole system. Within this
structure, orientational information spread locally and rapidly from site to
site. This is in contrast to the Vicsek model where information
spreads over large distances through the motion of dense groups of
particles (``flocks''). These results shows that even a small repulsive term can lead to fundamentally different
behavior of the system. Both, the
aligning as well as the disaligning part of the interaction can play
an important role
for the large-scale coherent motion of the particles. There is an abundance of systems that
are modeled with repulsive terms in
nature~\cite{Huth1992,Couzin2003,Gregoire2003,Chate2008a} (e.g. many particles avoid
collisions), in which this effect could be relevant.

Since information transfer in the grid-like and the flocking regimes
are fundamentally different, it is not evident that the phase
transition from order to disorder is of the same type in both
regimes. For instance, it is not clear, whether the phase transition
is continuous or discontinuous, or whether one has true long range
order in the grid-like regime: Since orientational information spreads
in a similar manner than in the XY-model, noise might destroy 
long-range order as predicted for 2d equilibrium systems through the Mermin-Wagner
theorem~\cite{mermin1966absence,toner2005hydrodynamics}. Answering
these questions would be the virtue of more extensive simulations than
were possible for this first study. Moreover, an analytical treatment
of this effect in terms of a kinetic
theory~\cite{Bertin2009,Ihle2011,Bialek2012} as well as a detailed
scaling analysis~\cite{Baglietto2008} are could help to clarify the
key control parameters in our system.

\ack We are grateful to helpful discussions with T. Vicsék at the
outset of this project.

\section*{References}
\bibliography{collectivemotion}
\end{document}